\documentclass[aps,prl,twocolumn,superscriptaddress,amsmath,amssymb,showpacs,floatfix]{revtex4}
\usepackage{graphicx}
\usepackage{bm}
\usepackage{color}

\begin{document}

\newcommand{\be}{\begin{equation}}
\newcommand{\ee}{\end{equation}}

\title{Comment on ``Nature of the Epidemic Threshold for the
Susceptible-Infected-Susceptible Dynamics in Networks''}

\author{Hyun Keun Lee}
\affiliation{School of Physics, Korea Institute for Advanced Study, Seoul 130-722, Korea}
\author{Pyoung-Seop Shim}
\affiliation{Department of Physics, University of Seoul, Seoul 130-743,
Korea}
\author{Jae Dong Noh}
\affiliation{Department of Physics, University of Seoul, Seoul 130-743,
Korea}
\affiliation{School of Physics, Korea Institute for Advanced Study,
Seoul 130-722, Korea}

\date{\today}


\maketitle
Recently, Bogu\~{n}\'{a} {\it et. al.}~\cite{Boguna13} claimed 
that the epidemic threshold of the susceptible-infected-susceptible~(SIS) 
model is zero on random scale-free~(SF) networks with the small-world~(SW)
property. They drew the conclusion by taking into account a 
long-range reinfection mechanism. 
In this Comment, we will show that such an effect is too
weak to guarantee an endemic phase with a {\em finite} fraction of infected 
nodes.
Consequently, the epidemic threshold can be finite in random SF networks.

There have been growing interests in the epidemic threshold $\lambda_c$ of
the SIS model~\cite{Castellano10,Goltsev12,Lee13}. 
A study based on the quenched mean-field~(QMF) theory combined with the 
linear stability analysis~(LSA) predicted that $\lambda_c$ vanishes 
in any complex networks with diverging maximum degree 
$k_{\rm max}$~\cite{Castellano10}. 
On the contrary, Ref.~\cite{Lee13} predicted the Griffiths phase 
in which the system evolves into a healthy state following a 
non-exponential slow relaxation dynamics. 
The Griffiths phase and a finite $\lambda_c$ were confirmed 
numerically in a certain class of SF networks that do not have the 
SW property~\cite{Lee13}. 
The localized nature of infection~\cite{Goltsev12} plays an important role
in the Griffiths phase.

Bogu\~n\'a {\it et. al.}~\cite{Boguna13} 
revisited the problem by taking into account a dynamic correlation
in a coarse-grained time scale. 
We start with Eq.~(3) of Ref.~\cite{Boguna13} for the probability 
$\rho_k$ that a node of degree $k$ is infected. 
It consists of a recovery rate $\bar\delta(k,\lambda) \simeq
e^{-a(\lambda)k}$ with $a(\lambda)\propto \lambda^2$ and an infection rate
$\bar\lambda(d_{kk'}) \simeq \lambda e^{-b(\lambda)(d_{kk'}-1)}$ where
$b(\lambda)=\ln(1+1/\lambda)$ and $d_{kk'}$ is the distance between nodes of
degree $k$ and $k'$. 
The key feature is the long-range infection, though the rate  
decays exponentially with the distance. The authors of Ref.~\cite{Boguna13}
investigated whether it could be relevant to overcome the localized nature
of infection especially in SW networks where $d_{kk'} =
\mathcal{O}(\ln N)$ with the number of nodes $N$.
They performed the LSA and
concluded that the epidemic threshold is zero in 
the random SF networks with the SW property.

Note that the LSA does not always yield the true epidemic 
threshold. We have already witnessed the failure of the LSA
applied to the QMF theory~\cite{Castellano10,Lee13}. One must assure that 
the infection is sustainable covering a {\em finite}
fraction of nodes above the threshold.

Consider the fraction of infected nodes 
that are allowed by Eq.~(3) of Ref.~\cite{Boguna13}. 
The rate of infection events at a node of degree $k$ is bounded by 
$\bar{\lambda}_k \leq N \sum_{k'}P(k') \bar{\lambda}(d_{kk'})$ where $P(k)$
is the degree distribution. 
In SW networks, one obtains
$\bar\lambda_k \leq \lambda N 
 \left(\frac{k}{N\langle k\rangle}\right)^{\bar{b}}
 \sum_{k'} k'^{\bar{b}} P(k')$ 
where $\bar{b}=b(\lambda)/\ln\kappa$ and $\kappa$ 
is the average branching factor~(see Eq.~(2) of Ref.~\cite{Boguna13}).
In SF networks~($P(k)\sim k^{-\gamma}$) with $\gamma>3$, it becomes
\begin{equation}
\bar\lambda_k \leq c \lambda \left( \frac{kk_{\rm max}}{N\langle
k\rangle}\right)^{\bar{b}} \leq 
c \lambda \left(\frac{k_{\rm
max}^2}{N\langle k\rangle}\right)^{\bar{b}} \propto \lambda
N^{-\left(\frac{\gamma-3}{\gamma-1}\right)\bar{b}} 
\end{equation}
with a constant $c$ and $k_{\rm max}\sim N^{1/(\gamma-1)}$.
A node remains infected when $\bar\lambda(k) >
\bar\delta(k)~(\simeq e^{-a(\lambda)k})$. 
Hence we obtain a necessary condition for infected nodes, given by 
\begin{equation}\label{kc}
k > k_c = \mathcal{O}(\ln N) \ .
\end{equation}
This condition can be easily understood by comparing the typical 
reinfection time scaling 
as $\ln \tau_{rein.} \sim \ln \bar\lambda^{-1} \sim \ln N$
and the typical recovery time scaling as 
$\ln \tau_{recov.} \sim \ln \bar\delta^{-1} \sim k$.
The reinfection wins when Eq.~(\ref{kc}) satisfies.

The condition in Eq.~(\ref{kc}) yields the upper bound for the density
$\rho$ of infected nodes: $\rho \lesssim \sum_{k>k_c} (\lambda k)P(k) 
\propto (\ln N)^{-(\gamma-2)}$. It vanishes in the $N\to \infty$ limit.
Therefore the system cannot be in the endemic phase even in SW networks
as long as the infection rates decay exponentially. 
This result invalidates the claim of Ref.~\cite{Boguna13}.

We add a few remarks: (i)~$\bar{\lambda}(d)$ was
evaluated assuming that the infection is only through 
a single path~\cite{Boguna13}. The effect of multiple paths can be
incorporated with an extra factor $\kappa^{d}$. 
The infection rate $\bar\lambda\sim e^{-(b-\ln\kappa)d}$
still decays exponentially for small $\lambda$, 
hence the scaling of $k_c$ is not affected. 
(ii)~Numerical simulation data were presented 
in Ref.~\cite{Boguna13} as an evidence for $\lambda_c=0$. Considered was 
a coverage defined as the fraction of distinct nodes 
{\em ever infected}. 
It is doubtful that the coverage could be a proper quantity 
to monitor the epidemic transition. 
The {\em instant} coverage is thought to be 
a proper one instead. A further numerical work in this direction is
necessary.

This work was supported by the Basic Science Research Program through the
NRF Grant No.~2013R1A2A2A05006776.

\end{document}